\documentclass[pra,twoside,showpacs,twocolumn]{revtex4}
\usepackage{times}
\usepackage{amssymb}
\usepackage{amsmath}
\usepackage{graphicx}
\usepackage{epsfig}
\usepackage{dcolumn}
\usepackage{color}
\usepackage{bm}

\begin{document}

\title{Quantum correlations and violation of Bell inequality induced by External Field in a two photon radiative cascade }

\author{Luqi Yuan and Sumanta Das}
\email{dsumanta31@yahoo.com}
\affiliation{ Institute for Quantum Science and Engineering $\&$ Department of Physics and Astronomy\\ Texas A\&M University, College Station, Texas, 77843}

\begin{abstract}
We study the polarization dependent second order correlation of a pair of photons emitted in a four level radiative cascade driven by an external field. It is found that the quantum correlations of the emitted photons, degraded by the energy splitting of the intermediate levels in the radiative cascade can be efficiently revived by a far detuned external field. The physics of this revival is linked to an induced stark shift  and the formation of dressed states in the system by the non-resonant external field. Further, we investigated the competition between the effect of the coherent external field and incoherent dephasing of the intermediate levels. We found that the degradation of quantum correlations due to the incoherent dephasing can be content for small dephasing with the external field. We also studied the non-locality of the correlations by evaluating the Bell's inequality in the linear polarization basis for the radiative cascade. We find that the Bell parameter decreases rapidly with increase in the intermediate level energy splitting or incoherent dephasing rate to the extent that there is no violation. However, the presence of an external field leads to  control over the degrading mechanisms and preservation of nonlocal correlation among the photons. This in turn can induce, violation of Bell's inequality in the radiative cascade for arbitrary intermediate level splitting and small incoherent dephasing.
\end{abstract}

\pacs{42.50.Ct, 42.50.Ar, 42.50.Hz, 03.65.Ud,}

\maketitle
\section{Introduction}
Quantum correlation of polarized photons emitted in a radiative cascade has been quite extensively studied since the early days of quantum mechanics \cite{Koc67, Cla69, Fre72}. Earlier studies in this regard were mainly focused on fundamental tests of quantum mechanics like the violation of Bell's inequality and existence of local hidden variable theories \cite{Bel65, Cla69, Fre72, Fry76, Asp82}. In recent times though such polarization correlation studies have gained importance in context to quantum information science particularly because of the entangled nature of the photon pairs. Note that many quantum information (QI) protocols based on quantum optics like quantum cryptography \cite{Eke91}, teleportation \cite{Urs04,Lan07}, efficient optical quantum computing \cite{Kni01} and long-distance quantum communication using quantum repeaters \cite{Bre98} requires an entangled photon pair per pulse. In general a large yield of such entangled photon pairs can be generated by nonlinear optical processes in bulk media \cite{Kie93} like the parametric down-conversion \cite{Kwi95}. However they are broadband, probabilistic and subject to Poissonian emission statistics leading to multipair emission \cite{Sca05}. In contrast a deterministic source of entangled photons would be able to suppress any multipair production and generate light pulses containing single photon pair with a high yield. This hence would render many of the above mentioned QI protocols much more efficient.

A suitable candidate for deterministic source of entangled photons turns out to be the radiative cascade emission from a single dipole, modeled as a four level system emitting a pair of photon in each excitation cycle \cite{Lar08, Das08}. Note that the origin of entangled photon pair in a four level radiative cascade is attributed to the plausibility of two indistinguishable decay pathways. In case of atomic dipoles one requires careful trapping and preparation to implement this. However it has been found that in quantum dots (QDs), such photon pair generation can be triggered readily \cite{Ben00}. The biexciton decay in a QD generates a photon pair which can ideally be entangled in their time-frequency \cite{Sim05} or polarization degrees of freedom \cite{Ste_n06,Ako06, Haf07}. In practice though, the path indistinguishability is hard to achieve in a radiative cascade. Even for QDs due to anisotropic electron-hole exchange interaction the degeneracy of the intermediate (\textit{excitonic}) state is lifted \cite{Gam96, Bay02} thereby destroying nonlocal properties of the emitted photons.

\begin{figure}[h]
\includegraphics[width=10.5 cm]{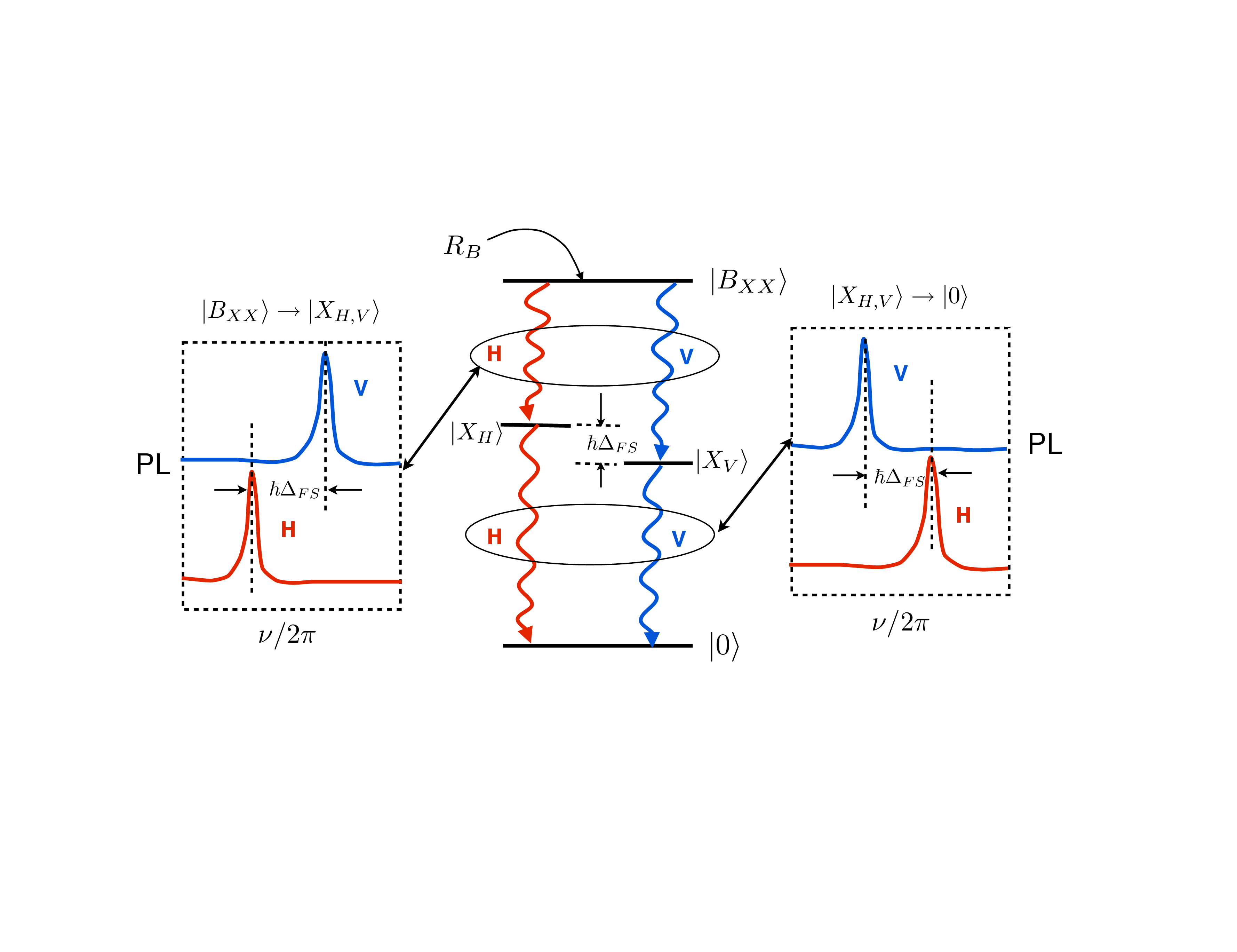}
\caption{(Color online) Schematic diagram of a general four level
radiative cascade.} \label{model1}
\end{figure}

The four level radiative cascade in general can be represented schematically by the Fig. (1).  We can see from the figure that due to the intermediate level splitting, the decay pathways are distinguishable and four distinct linearly polarized transition contribute to the emission spectrum. Entanglement of the polarized photons is then washed out given the fact that one can easily get which path information via energy consideration of the emitted photons \cite{San02}. This has hence lead to numerous investigations, particularly in QDs to find ways of reducing the intermediate state energy splitting within the radiative linewidth of the intermediate levels. \cite{Haf07, Seg05, Ger07,Jun08,Ste06,Ako06}. Note that, in a recent work a.c. Stark effect was used in QDs to reduced the intermediate level splitting within linewidth of the levels\cite{Mul09}. However it is worth mentioning here that even if the intermediate level splitting is cancelled in QDs, there are other processes that can have degradable effect on quantum correlations and polarization entanglement. For example dephasing interactions with the solid-state environment through collisions with phonons and electrostatic interactions with fluctuating charges around the dipoles \cite{Ber06} may also degrade the the polarization entanglement. Moreover, any incoherent mechanisms inducing a population exchange between the excitonic levels  such as transitions through the dark states or spin-flip processes may deteriorate the visibility of entanglement \cite{Lar08, Das08}.

We in this paper, theoretically investigate the effect of a strong external coherent drive on the intermediate level splitting in a four level radiative cascade and thereby on the second order quantum correlations of the emitted photons. We also evaluate the generalized form of Bell's inequality namely the Clauser-Horne-Shimony-Holt (CHSH) inequality \cite{Cla69, Asp82} in presence of the external coherent field and study how it can induce the generation of entanglement among the correlated photons. We restrict our analysis mainly to the rectilinear (H, V) polarization basis. We further consider other incoherent processes like the transfer of population among the intermediate states and investigate the effect of interplay of coherent field with this process on the quantum correlations and CHSH inequality.

\section{Model and Dynamical Evolution}
\subsection{Model of a radiative cascade}
\begin{figure}[!h]
\includegraphics[width=10 cm]{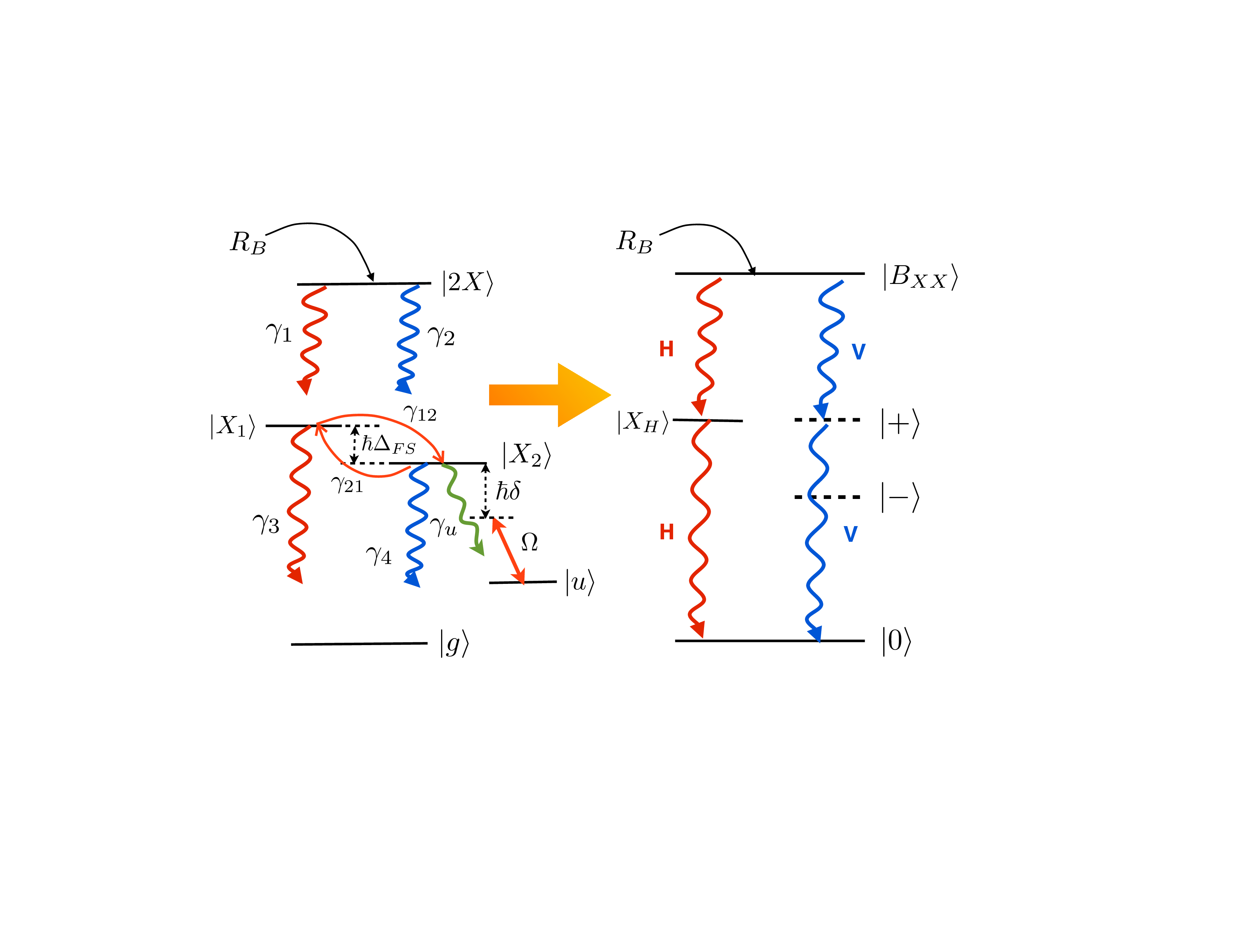}
\caption{(Color online) Energy level diagram of the five-level
cascade system. Here $\Delta_{FS}$ is the energy-level separation
of the intermediate states and $\gamma$'s are the spontaneous
emission rates. $\gamma_{12}$ and $\gamma_{21}$ are the incoherent
dephasing rates of the intermediate states. An external
non-resonant drive $\Omega$ couples one of the intermediate levels
$|X_{2}\rangle$ to an auxiliary level $|u\rangle$ with a detuning
$\delta$.} \label{model2}
\end{figure}

We consider a four level system undergoing a cascade emission as our primary model of study. In addition we also have an auxiliary level ($|u\rangle$) which is dipole allowed to $|X_{1}\rangle$ and $|X_{2}\rangle$. The importance of this auxlliary level in governing the dynamics of the four level cascade will be discussed later. The decay paths as shown schematically in Fig. (\ref{model2}) involves two radiative transitions, one from an upper level $|2X\rangle$ to the intermediate states $|X_{1}\rangle$ or $|X_{2}\rangle$ and the other from this intermediate states to the ground state $|g\rangle$. The energy-level splitting of the intermediate states is given by  $\Delta_{FS}$. Note that the basis states $\{|2X\rangle, |X_{1}\rangle, |X_{2}\rangle, |g\rangle\}$ of our model can correspond to the eigenbasis of a any dipole(like a QD or atom) under going a cascade decay. It is well known that radiative decay from the excited state in this basis generates collinearly polarized photons with two orthogonal linear polarizations $H$ (horizontal) and $V$ (vertical). Ideally if there is no splitting $\Delta_{FS} = 0$, the intermediate states are degenerate and the decay paths becomes indistinguishable. In this case the four level system relaxes, generating the maximally entangled two photon state
\begin{equation}\label{1}
|\Psi\rangle = \frac{1}{\sqrt{2}}\left[|H_{1}H_{2}\rangle+|V_{1}V_{2}\rangle\right],
\end{equation}
However, in practical situations like a QD cascade or an atomic cascade usually the splitting is nonzero and thus the intermediate levels are non-degenerate. Moreover population relaxation among the intermediate states can also occur for example from spin flipping processes. Thus in reality any polarization correlation or entanglement of the emitted photon pairs in such radiative cascade becomes crucially dependent on the degree of degeneracy and dynamics of these intermediate states. In this work we propose the use of external field to manipulate the dynamics of these intermediate states and thereby achieve control on the generation of polarization correlations of the emitted photons. For this purpose, we use an auxiliary state $|u\rangle$ in our model and couple it to one of the intermediate level  say $|X_{2}\rangle$ with a strong non-resonant coherent drive. The basic idea is to exploit the concept of a.c. Stark shift of the intermediate level $|X_{2}\rangle$ to nullify the level splitting $\Delta_{FS}$ and recover the indistinguishability of the decay paths. Since the Stark shift depends on the Rabi frequency of the transition $|X_{2}\rangle\rightarrow |u\rangle$ and the detuning $\delta$ (see Fig. \ref{model2}) of the drive it can in principle be made equal to $\Delta_{FS}$ even for arbitrary large value of $\Delta_{FS}$. In the next section we develop the theoretical framework necessary to study the dynamics of the four level cascade in presence of the auxiliary level.

\subsection{Dyamics of cascade emission}

To understand the dynamics of our system we consider a density matrix formalism ( given the open nature of the four level cascade ) and write down the corresponding master equation for the density operator $\rho$ as,
\begin{equation}\label{2}
\frac{\partial \rho}{\partial t} = -\frac{i}{\hbar}\left [\mathcal{H}, \rho\right ]+\mathcal{L}\rho.
\end{equation}
Here $\mathcal{H}$ is the Hamiltonian of the system comprising of the free energy term $\mathcal{H}_{0}$ and the interaction term $\mathcal{H}_{I}$,
\begin{eqnarray}\label{3}
\mathcal{H}_{0}  & = &\sum_{k}\hbar\omega_{k}|k\rangle\langle k|\nonumber\\
\mathcal{H}_{I} & =& - \hbar \left( {\Omega e^{i\delta t} \left| {X _2 }
\right\rangle \left\langle u \right| + \Omega ^ *  e^{ - i\delta
t} \left| u \right\rangle \left\langle {X _2 } \right|}
\right).
\end{eqnarray}
Here $\Omega = \vec{d}_{X_{2}u}\cdot\vec{\mathcal{E}}/ \hbar $ is the Rabi frequency of the transition and $\delta$: the detuning given by $\delta  = \omega_{X_{2}u} - \nu_{L}$, where $\nu_{L}$ is the frequency of the incident coherent drive. The relaxation processes like radiative decay and other incoherent mechanism present in the system that lead to decoherence is incorporated in the dynamics by the Lindblad operator $\mathcal{L}\rho$ of Eq. (\ref{2}) and is given by,
\begin{eqnarray}\label{4}
\mathcal{L}\rho & = & \mathcal{L}_{r}\rho+\mathcal{L}_{d}\rho\\
\mathcal {L}_{r} \rho & = &  - \sum_{i= 1}^{5} {\frac{{\gamma _i
}}{2}\left( {\sigma _ + ^i \sigma _ - ^i \rho  - 2\sigma _ - ^i
\rho \sigma _ + ^i  + \rho \sigma _ + ^i \sigma _ - ^i } \right)},\\
\mathcal {L}_{d} \rho & = &  -\frac{\gamma _{12}}{2}\left( {\sigma _ +^{\alpha} \sigma _ - ^{\alpha}\rho  - 2\sigma _ - ^{\alpha}\rho \sigma _ + ^{\alpha}  + \rho \sigma _ + ^{\alpha} \sigma _ - ^{\alpha} } \right)\nonumber\\
& & -\frac{\gamma _{21}}{2}\left( {\sigma _ +^{\beta} \sigma _ - ^{\beta}\rho  - 2\sigma _ - ^{\beta}\rho \sigma _ + ^{\beta}  + \rho \sigma _ + ^{\beta} \sigma _ - ^{\beta} } \right).
\end{eqnarray}
Here the operators $\sigma_{+}$ = $\sigma_{-}^{\dagger}$ are the atomic lowering and raising operators  defined as : $\sigma_{+}^{1} = |2X\rangle\langle X_{1}|, ~\sigma_{+}^{2} = |2X\rangle\langle X_{2}|, ~\sigma_{+}^{3} = |X_{1}\rangle\langle g|, ~\sigma_{+}^{4} = |X_{2}\rangle\langle g|,~ \sigma_{+}^{5} = |X_{2}\rangle\langle u|,~ \sigma_{+}^{\alpha} = |X_{1}\rangle\langle X_{2}|$ and  $\sigma_{+}^{\beta} = |X_{2}\rangle\langle X_{1}|$. The decays in the system (see Fig. 2 ) are given as follows : $\gamma = (\gamma_{1}+\gamma_{2})$ and $\gamma_{3}$ ($\gamma_{4}$) are the radiative decay rate of the excited state $|2X\rangle$ and the state $|X_{1}\rangle$ ($|X_{2}\rangle$) respectively. The radiative decay rate on the transition $|X_{2}\rangle \rightarrow |u\rangle$ is given by $\gamma_{5} = \gamma_{u}$. Further, $\gamma_{12} (\gamma_{21})$ is the incoherent dephasing rate of the state $|X_{1}\rangle (|X_{2}\rangle)$ which corresponds to the rate of population relaxation between them. It is worth mentioning here that the incoherent dephasing considered by us is different from the pure dephasing in qubits. We here consider decoherence arising from population relaxation of the intermediate state as well as decay of the coherence among the levels whereas in pure dephasing environment only the coherence decays. As such the Lindbald operator to model these decoherence phenomenas (incoherent dephasing) take the form $\mathcal{L}_{d}$ \cite{Das08}[check for example Eq.$[(A-7), (A-12), (A-13)]$ of appendix A]. In accordance with the above framework the time evolution of our system reduces to a set of differential equations of the form
\begin{equation}\label{5}
\frac{d\rho}{dt} = M \rho,
\end{equation}
where $M$ a $25\times25$ sparse square matrix. An elaborate form
of Eq. (\ref{5}) containing the detail of the time dependence of
the density matrix elements is provided in the appendix A. To
understand the effect of external field induced manipulation of
the intermediate level splitting and incoherent dephasing on
the dynamics of photon emission, we next study the two-time second-order correlations.

\section{Quantum correlation of photon pairs}
The two time second order correlation is an experimentally measurable quantity being related to the two time intensity correlations of the emitted photons.  Further it also reflects the influence of the atomic properties on the statistics of the emitted photons. The two time second order correlation is defined by \cite{Das08},
\begin{eqnarray}
\label{6}
\langle I I \rangle & = &\langle\hat{\epsilon}^{\ast}_{(\theta_{1},\phi_{1})}\cdot \vec{E}^{-}(\vec{r},t)\hat{\epsilon}^{\ast}_{(\theta_{2},\phi_{2})}\cdot \vec{E}^{-}(\vec{r},t+\tau)\nonumber\\
& &:\hat{\epsilon}_{(\theta_{2},\phi_{2})}\cdot \vec{E}^{+}(\vec{r},t+\tau)\hat{\epsilon}_{(\theta_{1},\phi_{1})}\cdot \vec{E}^{+}(\vec{r},t)\rangle.\nonumber\\
\end{eqnarray}
where $\langle I I \rangle$ stands for the two time polarization angle dependent intensity-intensity correlation $\langle I_{(\theta_{2},\phi_{2})}(\vec{r},t+\tau)I_{(\theta_{1},\phi_{1})}(\vec{r},t)\rangle$.
Note that in the above definition, $E^{+}(\vec{r},t) (E^{-}(\vec{r},t))$ is the positive (negative) frequency part of the quantized electric field operator at a point $\vec{r}$ in the far-field zone. The electric field operator for our model of the radiative cascade is given by,
\begin{eqnarray}
\label{7}
\vec{E}^{(+)}(\vec{r},t) & = &\vec{E}^{(+)}_{0}(\vec{r},t) -\left(\frac{\omega_{0}}{c}\right)\frac{1}{r}(\left[\hat{n}\times(\hat{n}\times\vec{d}_{X_{1}2X})\right]|X_{1}\rangle\langle 2X|_{t}\nonumber\\
& + &\left[\hat{n}\times(\hat{n}\times\vec{d}_{X_{2}2X})\right]|X_{2}\rangle\langle 2X|_{t}\nonumber\\
& + &\left[\hat{n}\times(\hat{n}\times\vec{d}_{g X_{1} })\right]|g\rangle\langle X_{1}|_{t}\nonumber\\
& + &\left[\hat{n}\times(\hat{n}\times\vec{d}_{g X_{2}})\right]|g\rangle\langle X_{2}|_{t}).
\end{eqnarray}
Further, $\hat{\epsilon}_{(\theta,\phi)}$ is the polarization unit vector of the measured radiation at the detector along any arbitrary direction given by $(\theta,\phi)$ and are related to the linear polarization unit vectors $\hat{\epsilon}_{H},\hat{\epsilon}_{V}$ (where H stands for horizontal and V for vertical) by \cite{Lar08,Das08},
\begin{equation}
\label{7a}
\left[\begin{array}{c} \hat{\epsilon}^{(1)}_{(\theta,\phi)} \\
\\
\hat{\epsilon}^{(2)}_{(\theta,\phi)}
\end{array}\right] =
\left[\begin{array}{cc} \cos\theta & e^{-i\phi}\sin\theta \\
\\
-e^{i\phi}\sin\theta & \cos\theta \end{array}\right]
\left[\begin{array}{c}\hat{\epsilon}_{H} \\
\\
\hat{\epsilon}_{V}\end{array}
\right]
\end{equation}
The polarization unit vectors satisfies the orthogonality relation
$(\hat{\epsilon}^{(i)}_{(\theta,\phi)}\cdot\hat{\epsilon}^{(j)\ast}_{(\theta,\phi)}
)= \delta_{ij}$. The above matrix relation can be understood as an unitary
transformation between a basis defined by the linear polarization
unit vectors and a basis defined by $\hat{\epsilon}^{(1)}$ and
$\hat{\epsilon}^{(2)}$. In experimental setup the angles $\theta, \phi$ would correspond to the orientation of the optic axis of a half/quarter wave plate to the direction of propagation of the emitted radiation. We will next write down the key expression for the $\langle I I\rangle$ and discuss the implication. For this purpose let us first make some simplified assumptions. Let us consider that both the levels $|X_{1}\rangle$ and $|X_{2}\rangle$ in Fig. 2 have the same incoherent
dephasing rates i.e. $\gamma_{12} = \gamma_{21} = \gamma_{d}$.
Further we assume that the radiative decay rates of the intermediate levels are also equal $(\gamma_{1} = \gamma_{2} = \gamma_{3} = \gamma_{4} = \gamma)$. Such assumptions are well justified as they do not influence the dynamics of the system significantly.  Moreover as we restrict our analysis to the linear polarization basis, we can set  $\phi_{1} = \phi_{2} = 0$. Hence under the above assumptions the two-time polarization angle dependent intensity-intensity correlation is found to be:
\begin{eqnarray}
\label{8}
 \left\langle {I I} \right\rangle  & = & \left( {\frac{{\omega _0 }}{c}} \right)^8 \frac{1}{{2r^4 }}D_1^2 D_2^2 \langle | 2X \rangle \langle 2X |_t \rangle \{ e^{ - \gamma \tau } \nonumber\\
 & +& \cos 2\theta _1 \cos 2\theta _2 ~ e^{ - (\gamma  + 2\gamma _{d})\tau } \nonumber\\
 &+&2 \sin 2\theta _1 \sin 2\theta _2 {\mathop{\rm Re}\nolimits} \left[ \mathcal{W}\left( \Omega, \delta, \tau  \right) \right]\},
\end{eqnarray}
where,
\begin{eqnarray}
\label{9}
\mathcal{W}(\Omega, \delta, \tau  ) & = & \exp\left [-\left\{\frac{3\Gamma}{4}+i(\Delta_{FS}+\delta/2)\right \} \tau\right ] \times\nonumber\\
& & \left[ \cos \left( \zeta \tau \right)+ \frac{\Gamma_{1}-i\delta/2 }{\zeta }\sin \left( \zeta \tau \right) \right]
\end{eqnarray}
and $\Gamma  =  (\gamma+\gamma_{d}+\gamma_{u}/3),~\Gamma_{1}  =  (\gamma+\gamma_{d}+\gamma_{u}),~\zeta = \sqrt{\Omega^{2}-(\Gamma_{1}/4-i\delta/2)^{2}}$. The simplified form of the second order correlation in Eq. (\ref{8}) has been derived to gain helpful insight, about the effect of interplay between the external field, the intermediate level splitting $\Delta_{FS}$ and the other incoherent processes on the polarization correlation of photon pairs. For detailed mathematical analysis leading to the generalized form of the two time intensity-intensity correlation, the reader is referred to appendix B of this paper.  One can clearly see from Eqns (\ref{8}, \ref{9}) that the second order correlation is profoundly influenced by the incoherent dephasing rate $\gamma_{d}$ as well as the excitonic level splitting $\Delta_{FS}$, the Rabi frequency $\Omega$ and detuning $\delta$. We next study the second order correlation of Eq. (\ref{8}) for different limits of the system parameters to understand their effect on the dynamics of photon pair correlation and generation. \\
\\
\textbf{Case I :} To understand the sole effect of the splitting $\Delta_{FS}$ we first consider $(\gamma_{d} = \Omega = \delta = 0)$. In the absence of any external field and for $\gamma >> \gamma_{u}$ the intensity -intensity correlation of Eq. (\ref{8}) reduces to the second order correlations measured in ref.\cite{Ste_n06,Ste06} and given by \cite{Das08},
\begin{eqnarray}
\label{10}
 \left\langle {I I} \right\rangle  & = & \left( {\frac{{\omega _0 }}{c}} \right)^8 \frac{1}{{2r^4 }}D_1^2 D_2^2 \langle | 2X \rangle \langle 2X |_t \rangle  e^{ - \gamma \tau }\{1+ \nonumber\\
 & & \cos 2\theta _1 \cos 2\theta _2 + \sin 2\theta _1 \sin 2\theta _2 \cos (\Delta_{FS} \tau)\}. \nonumber\\
\end{eqnarray}
Further, note that for small $\Delta_{FS}$ the above expression for intensity-intensity correlation is equivalent to that proposed by Freedman and Clauser \cite{Fre72} and later measured by Aspect and co-workers \cite{Asp82}. \\
\\
\textbf{Case II :}  We next study the effect of a strong resonant $(\delta = 0)$ external field on $\Delta_{FS}$ and thereby on the second order correlation. For this purpose we let $\gamma_{d} = 0$ and neglect $\gamma_{u}$ under the assumption that $\gamma >> \gamma_{u}$. Further we also consider that $\Omega >> $ all the decay rates in the system. With these assumptions,  Eqn. (\ref{8}) reduces to
\begin{eqnarray}
\label{11}
 \left\langle {I I} \right\rangle &  = & \left( {\frac{{\omega _0 }}{c}} \right)^8 \frac{1}{{2r^4 }}D_1^2 D_2^2 \langle | 2X \rangle \langle 2X |_t \rangle e^{ - \gamma \tau }\{1+ \nonumber\\
 & & \cos 2\theta _1 \cos 2\theta _2+  \sin 2\theta _1 \sin 2\theta _2 e^{ \gamma \tau/4 }\nonumber\\
 &\times&[\cos\{ (\Delta_{FS}+\Omega) \tau\}+\cos\{ (\Delta_{FS}-\Omega) \tau\}]\}.\nonumber\\
\end{eqnarray}
Thus we see that in presence of a resonant external field, beats result in the system with a frequency equal to the Rabi frequency $\Omega$. Now for $\Delta_{FS}=\Omega$ we find,
\begin{eqnarray}
\label{11a}
\langle I I\rangle & \propto & e^{ - \gamma \tau }[1+ \cos 2\theta _1 \cos 2\theta _2]\nonumber\\
& + & \sin 2\theta _1 \sin 2\theta _2 e^{ -3\gamma \tau/4 }\{1+\cos \left (2\Omega\tau\right)\}.
\end{eqnarray}
It can be clearly seen that the second order correlation oscillates with twice the Rabi frequency. Next to analytically study the behavior of this correlation let us consider some approximations. We see from the above expression that when $\Delta_{FS} = \Omega >> \gamma, (2 \Omega \tau)$ is very large and hence $\langle\cos(2 \Omega \tau)\rangle_{\tau}$ can be neglected. Here $\langle .... \rangle_{\tau}$ stands for time average.Thus under this approximation the timed average second order correlation becomes independent of intermediate level splitting or Rabi frequency and depends only on the decay rates and polarization angle of the detectors. The reason of considering the time average of the second order correlation and this approximation will become apparent in the next section.\\
\\
\textbf{Case III :} In the case of a far detuned $(\delta \geq \Omega )$ strong external field with $\gamma_{d} = 0$ and neglecting  $\gamma_{u}$ the second order correlation becomes
\begin{eqnarray}
\label{12}
 \left\langle {I I} \right\rangle &  = & \left( {\frac{{\omega _0 }}{c}} \right)^8 \frac{1}{{2r^4 }}D_1^2 D_2^2 \langle | 2X \rangle \langle 2X |_t \rangle e^{ - \gamma \tau }\{1+ \nonumber\\
 & & \cos 2\theta _1 \cos 2\theta _2+  \sin 2\theta _1 \sin 2\theta _2~ e^{ \gamma \tau/4 }\nonumber\\
 &\times&[\cos\{ (\Delta_{FS}+\Omega_{+}) \tau\}+\cos\{ (\Delta_{FS}-\Omega_{-}) \tau\}]\},\nonumber\\
 \end{eqnarray}
 where $\Omega_{\pm} = 1/2\sqrt{\delta^{2}+4\Omega^{2}}\pm\delta/2$ and we have considered $\Omega, \delta >> \gamma$. Thus here again we see that in presence of a resonant external field, beats result in the system with a frequency $\Omega_{+}+\Omega_{-} = \sqrt{\delta^{2}+4\Omega^{2}}$. Note that in this case when $\Delta_{FS} = \Omega_{-}$ the intensity-intensity correlation becomes
\begin{eqnarray}
\label{12a}
\left\langle {I I} \right\rangle &  \propto & e^{ - \gamma \tau }[1+ \cos 2\theta _1 \cos 2\theta _2 ]\nonumber\\
 & + &  \sin 2\theta _1 \sin 2\theta _2~ e^{ -3\gamma \tau/4 }[1+\cos\{(\sqrt{\delta^{2}+4\Omega^{2}}) \tau\}],\nonumber\\
 \end{eqnarray}
It is interesting to compare the time dependent cosine modulation term of Eq. (\ref{11a}) and (\ref{12a}). We see that when we have a detuned external field, for any $\Delta_{FS}$ the argument of the modulation will be ${(2 \Delta_{FS} +\delta) \tau}$. This is obviously greater than the modulation of $(2 \Delta_{FS} \tau)$ that we get from Eq. (\ref{11a}) for the same $\Delta_{FS}$. Thus in this case our earlier approximation regarding dropping the highly oscillating cosine will work much better. Thus we anticipate that our scheme for a strong external field with large detuning will give optimal results for suppressing the effect of intermediate level splitting on the two photon correlations.   \\

\textbf{Case IV :} Finally we will incorporate the incoherent dephasing of the intermediate states $|X_{1,2}\rangle$ in our analysis and study its effect on the quantum correlation of photons pairs. In this case we consider all the dynamical parameters with only two reasonable assumption : $\Omega, \delta >>$ all the decay rates in the system and that  $\gamma >> \gamma_{u}$. The second order correlation then becomes,
\begin{eqnarray}
\label{13}
 \left\langle {I I} \right\rangle &  = & \left( {\frac{{\omega _0 }}{c}} \right)^8 \frac{1}{{2r^4 }}D_1^2 D_2^2 \langle | 2X \rangle \langle 2X |_t \rangle e^{ - \gamma \tau }\{1+ \nonumber\\
 & & \cos 2\theta _1 \cos 2\theta _2~e^{-2\gamma_{d}}+  \sin 2\theta _1 \sin 2\theta _2~ e^{ [\gamma-3\gamma_{d}]\tau/4 }\nonumber\\
 &\times&[\cos\{ (\Delta_{FS}+\Omega_{+}) \tau\}+\cos\{ (\Delta_{FS}-\Omega_{-}) \tau\}]\},\nonumber\\
 \end{eqnarray}
Thus we see that in presence of the incoherent dephasing all the polarization angle dependent terms in the correlation gets an additional decay which will crucially regulate the quality of correlation among the photons. The approximation we considered in the earlier cases also hold in this case but they does not effect the dephasing rate of its influence. In the next section we define a quantitative measure of polarization correlation among the photons and discuss the effect of the external field, excitonic splitting and incoherent dephasing on polarization correlation and entanglement generation among the photons.

\section{Degree of Polarization Correlation}
In this section we study a quantity: the time averaged degree of correlation $C_{\mu}$ in a basis defined by the polarization of the emitted photons. Note that $C_{\mu}$ has been studied extensively in context to polarization entanglement \cite{Ste_n06,Ste06,You09} and is defined in the literature as,
\begin{equation}
\label{14}
C_{\mu} = \frac{\langle I_{\mu}I_{\mu}\rangle -\langle I_{\mu}I_{\mu^{\prime}}\rangle}{\langle I_{\mu}I_{\mu}\rangle +\langle I_{\mu}I_{\mu^{\prime}}\rangle}
\end{equation}
where  $\mu, \mu^{\prime}$ stands for mutually orthogonal polarization basis like $\{H, V\}$ or $\{D, D^{\prime}\}$. The degree of correlation varies between $+1$ and $-1$, where $+1$ represent perfect correlation ($-1$ for anti-correlation) and 0 represent no polarization correlation. We next investigate the effect of intermediate level splitting $\Delta_{FS}$, the external field $\Omega$, detuning $\delta$ and the incoherent dephasing rate $\gamma_{d}$ on the degree of correlation. For this purpose we consider a time average of Eq. (\ref{14})
such that $C_{\mu}$ is solely dependent on the polarization angles for some particular values of $\Delta_{FS}, \Omega,\delta$ and $\gamma_{d}$. This thus helps to understand the effect of interplay among several dynamical parameters of the system on the quantum correlation of the polarized photons.

\subsection{Effect of the external field on the quantum correlation among photons in presence of intermediate level splitting}

In this sub-section we numerically study (\textit{without any approximation on Eq. \ref{8}}) the behavior of the degree of correlation $C_{\mu}$ averaged over time as a function of basis angle, for non zero intermediate-level splitting $\Delta_{FS}$ and different external fields strength. We neglect any incoherent dephasing mechanism (put $\gamma_{d} = 0$) for present, to keep our discussion simple. Effect of such incoherent dephasing in presence of the intermediate level splitting and external field will be considered in the next section.

\begin{figure}[!h]
\includegraphics[width=8.0cm]{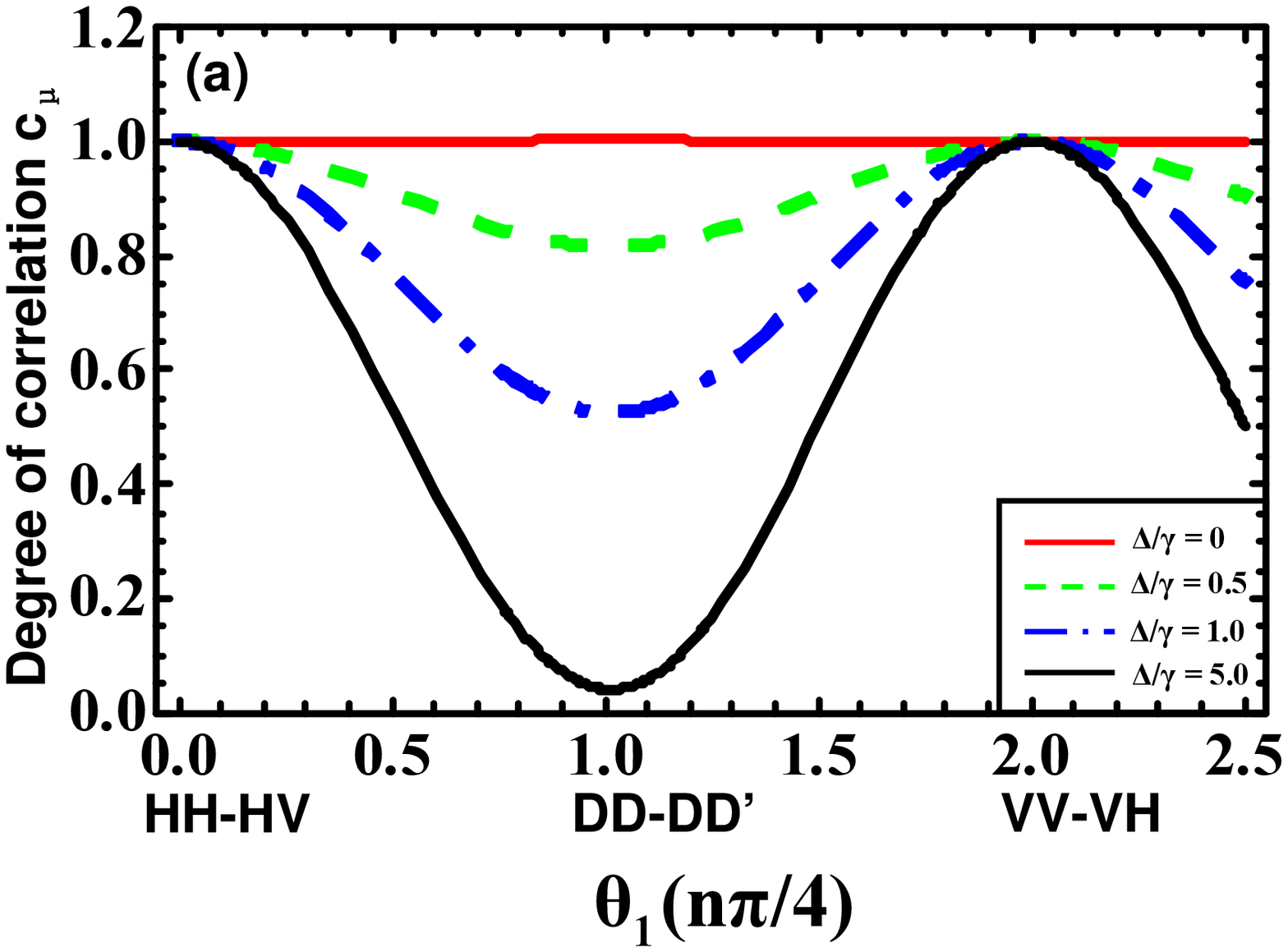}
\includegraphics[width=8.0cm]{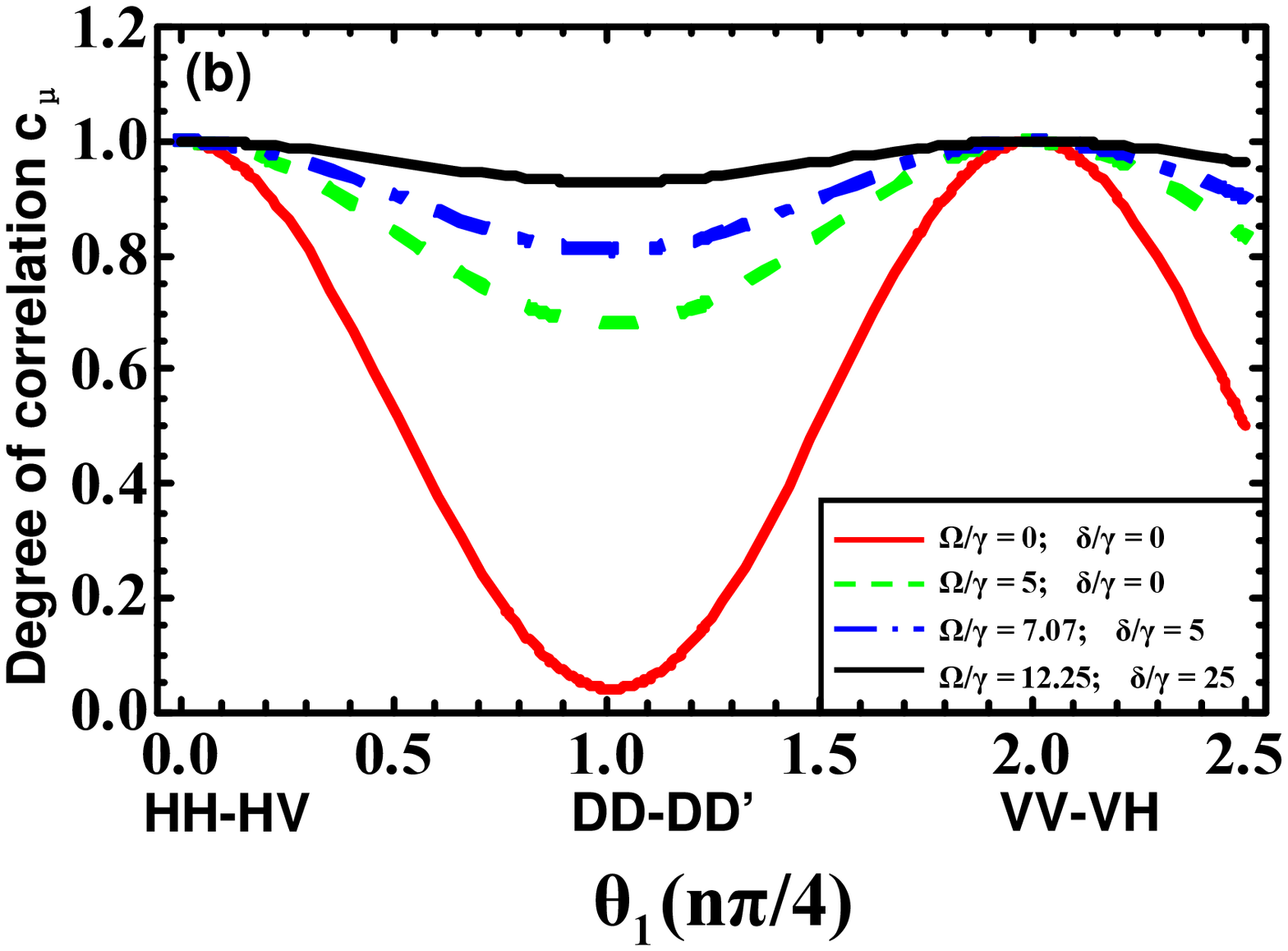}
\includegraphics[width=8.0cm]{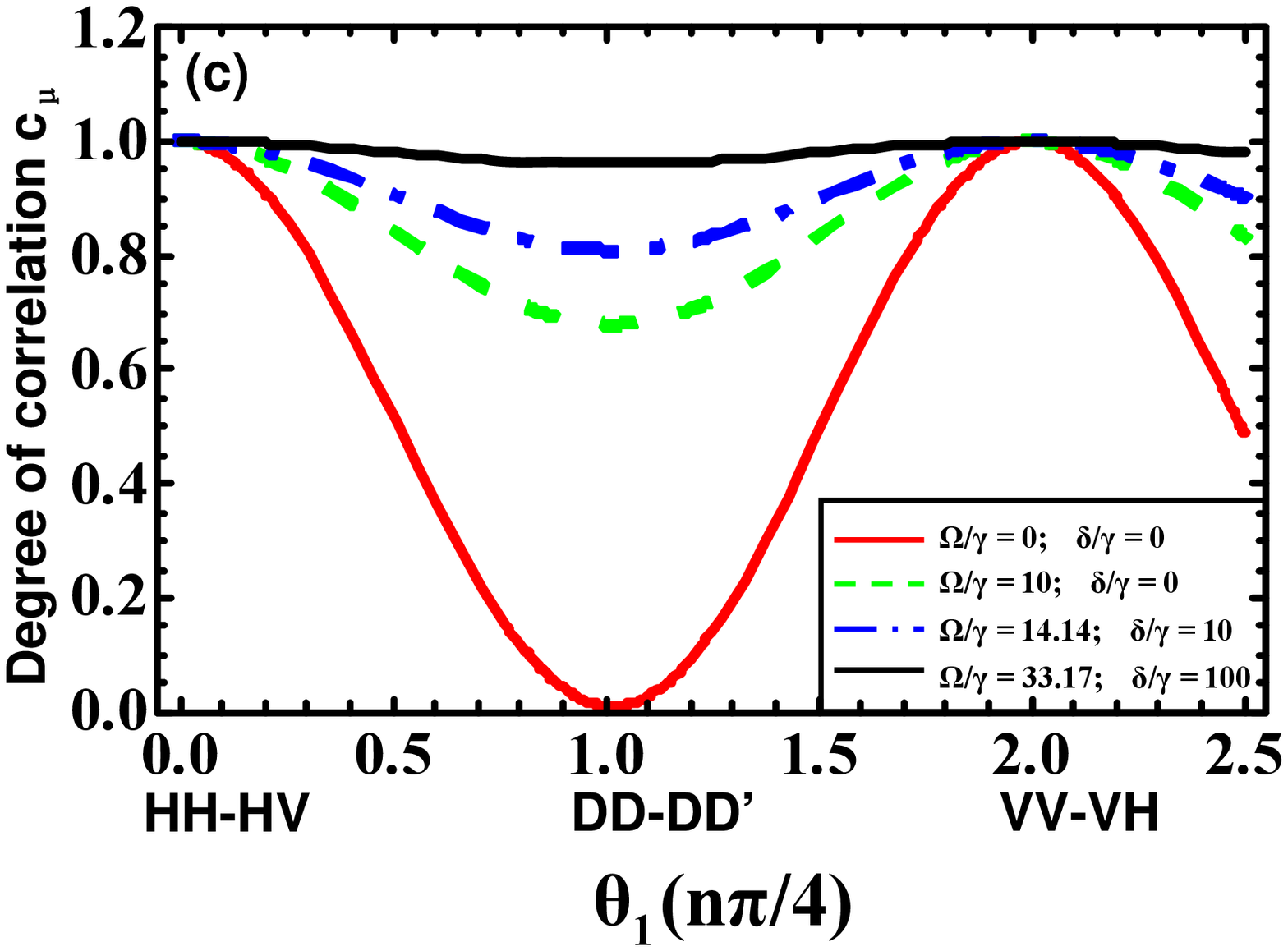}
\caption{(Color online) Degrees of correlation averaged over time
as a function of basis angle for (a) different intermediate-level
splitting without external field and with different drive field
for (b) $\Delta_{FS} /\gamma = 5$ and (c) $\Delta_{FS} / \gamma =
10$. The drive field is given by $\Omega / \gamma = \sqrt
{\left(\Delta_{FS} ^2 + \Delta_{FS} \delta \right) /  \gamma^2}
$.} \label{AvCorrNoDephase}
\end{figure}

In Fig. (\ref{AvCorrNoDephase}) we show the behavior of time averaged $C_{\mu}$ as a function of the basis angle. We find that for no intermediate-level splitting ($\Delta_{FS} = 0$), there is perfect quantum correlation and thereby entanglement among the photons in the polarization basis ( for a detail discussion in this regard see for example ref \cite{Ste_n06,Das08} ). This is reflected in the figure by the fact that the degree of correlation $C_{\mu}$ does not change (solid straight line of Fig. \ref{AvCorrNoDephase}a) when the observation is made in different polarization bases ($HV-DD-VH$).
However we find that as the intermediate level splitting increases the quantum correlation among the photons in the polarization basis is degraded. This is depicted by the oscillatory behavior of $C_{\mu}$ in  Fig.(\ref{AvCorrNoDephase}a) as the observation of photons are made in different polarization basis. Finally we find (solid oscillatory curve in Fig. \ref{AvCorrNoDephase}a) that the quantum correlation among the photons are completely lost for large intermediate level splitting. Note that similar behavior was reported earlier in the study of such polarization correlation in Ref.~\cite{Das08}. Now that we have analyzed the dependence of degree of correlation and thereby the quantumness of two photon correlation on the intermediate level splitting we focus on the key aspect of this paper.

We now consider the effect of an external coherent field on the degree of correlation $C_{\mu}$ when $\Delta_{FS} \neq 0$. In Fig. (\ref{AvCorrNoDephase}b) and (\ref{AvCorrNoDephase}c) we show that $C_{\mu}$ is dramatically altered in presence of the field. For example we see from Fig. (\ref{AvCorrNoDephase}b) that for $\Delta_{FS} = \Omega = 5\gamma$ the degree of correlation of the photons in the diagonal basis is enhanced by almost $75\%$. This can be further enhanced to the extent of achieving almost perfect quantum correlations among the photons with higher field strength (around $\Omega \sim 2.5 \times \Delta_{FS} = 12.25\gamma $) and strong detuning ($\delta = 5\times\Delta_{FS}= 25\gamma$ in this case). Thus we find that our scheme becomes more effective with suitable off-resonant external field and larger field strengths. Moreover this also suggest that our approximation of dropping the cosine term in the analytical discussion of the quantum correlation is more appropriate for the case (III).  Note that parameters used in our simulation are well within reach of experiments in practically realizable systems \cite{Mul09,You09}.
In Fig. (\ref{AvCorrNoDephase}c) we show the effect of external field on the degree of correlation for a representative value of the intermediate level splitting $\Delta_{FS} = 10\gamma$. Our simulations predict even better result for this case with large detuning (almost $10\times\Delta_{FS}$) and field strength ($3.5\times\Delta_{FS}$).

The results of our simulations can be understood under dressed state basis. The
external field between level $X_2$ and level $u$ will split the
excitonic level into two eigenstates $|+\rangle$ and $|-\rangle$
as shown in Fig. (\ref{model2}). When the field satisfies the
condition that $\Omega = \sqrt {\Delta_{FS} ^2  + \Delta_{FS} \delta }$, the
eigenstate $|+\rangle$ has the same energy as the other excitonic
level $X_1$. There is no energy splitting between $|X_1\rangle$
and $|+\rangle$, so the polarization correlation in the diagonal
basis can be revived as shown in Fig. (\ref{AvCorrNoDephase}).
However, another eigenstate $|-\rangle$ also affects the emission
of photon from the biexciton level. This effect makes the degree
of the correlation not perfectly recover back to 1. The energy
difference between two eigenstates is $\sqrt {\delta ^2  + 4\Omega
^2 }$. Therefore, larger detuning $\delta$ can make this
disturbing effect by eigenstate $|-\rangle$ be weak enough until
it can be approximately neglected and the degree of correlation in
the diagonal basis is then enhanced to almost 1.

Our numerical finding hence suggests that external control fields
can efficiently revive degraded quantum correlation among the
photons for quite large values of intermediate level splitting
also. We would like to emphasize here that our theoretical model
and methods thus predicts a practically plausible \textit{control
knob} ($\Omega, \delta$) to regulate the intrinsic parameter of
the system ($\Delta_{FS}$) which is otherwise difficult to achieve
in such radiative cascade systems
\cite{Haf07,Seg05,Ste06,Ger07,Jun08}. Of course, our results have certain limitation on the
choice of $\Omega$ and $\delta$. It is important to note that this
parameters can in principle be arbitrary, although off-resonant
external field with moderately large power but large detuning is
not realistic in practice because it may not be in the absorption
width.

\subsection{Effect of the external field on the quantum correlation among photons in presence of Incoherent processes}

\begin{figure}[!h]
\includegraphics[width=8.0cm]{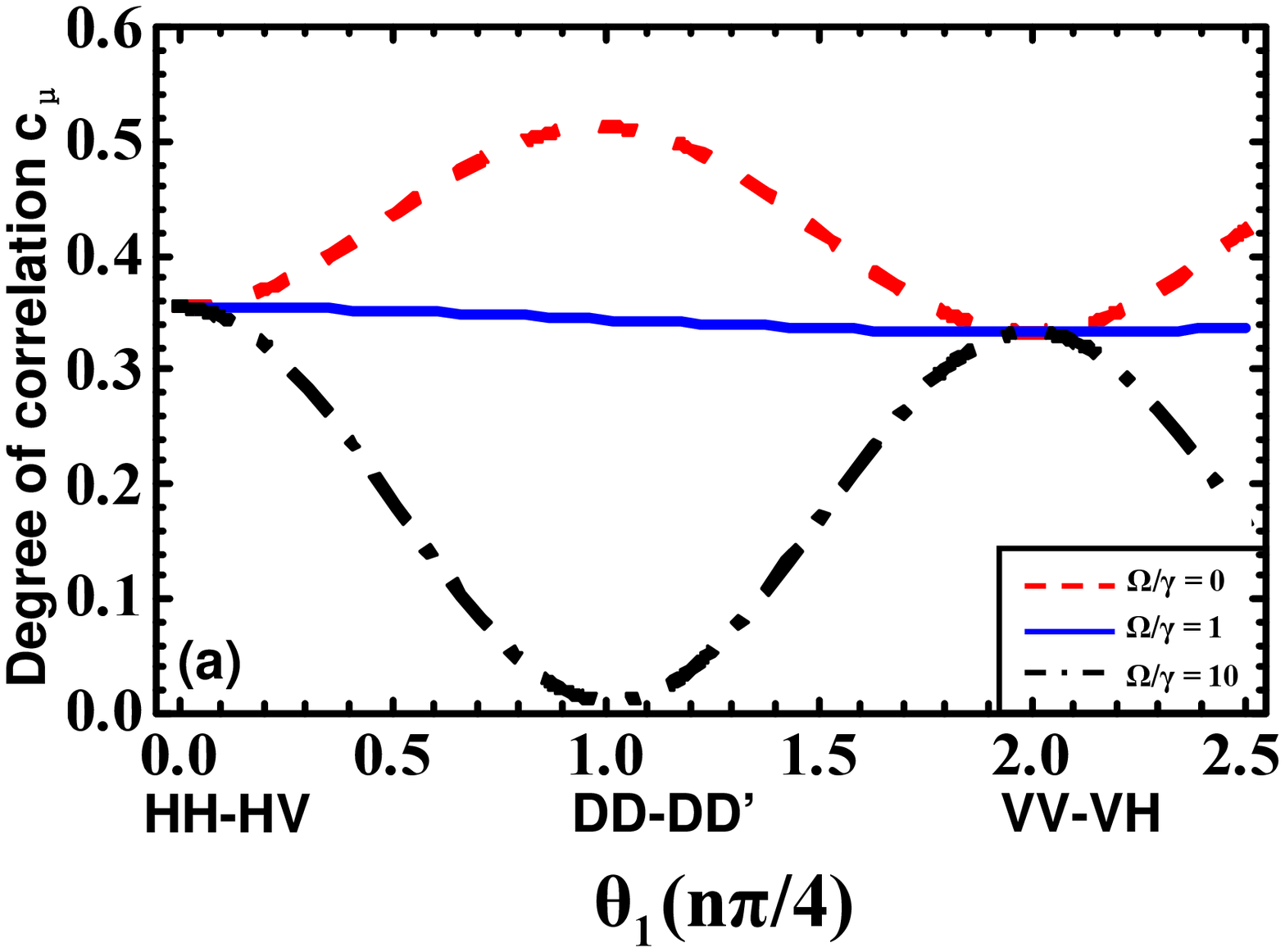}
\includegraphics[width=8.0cm]{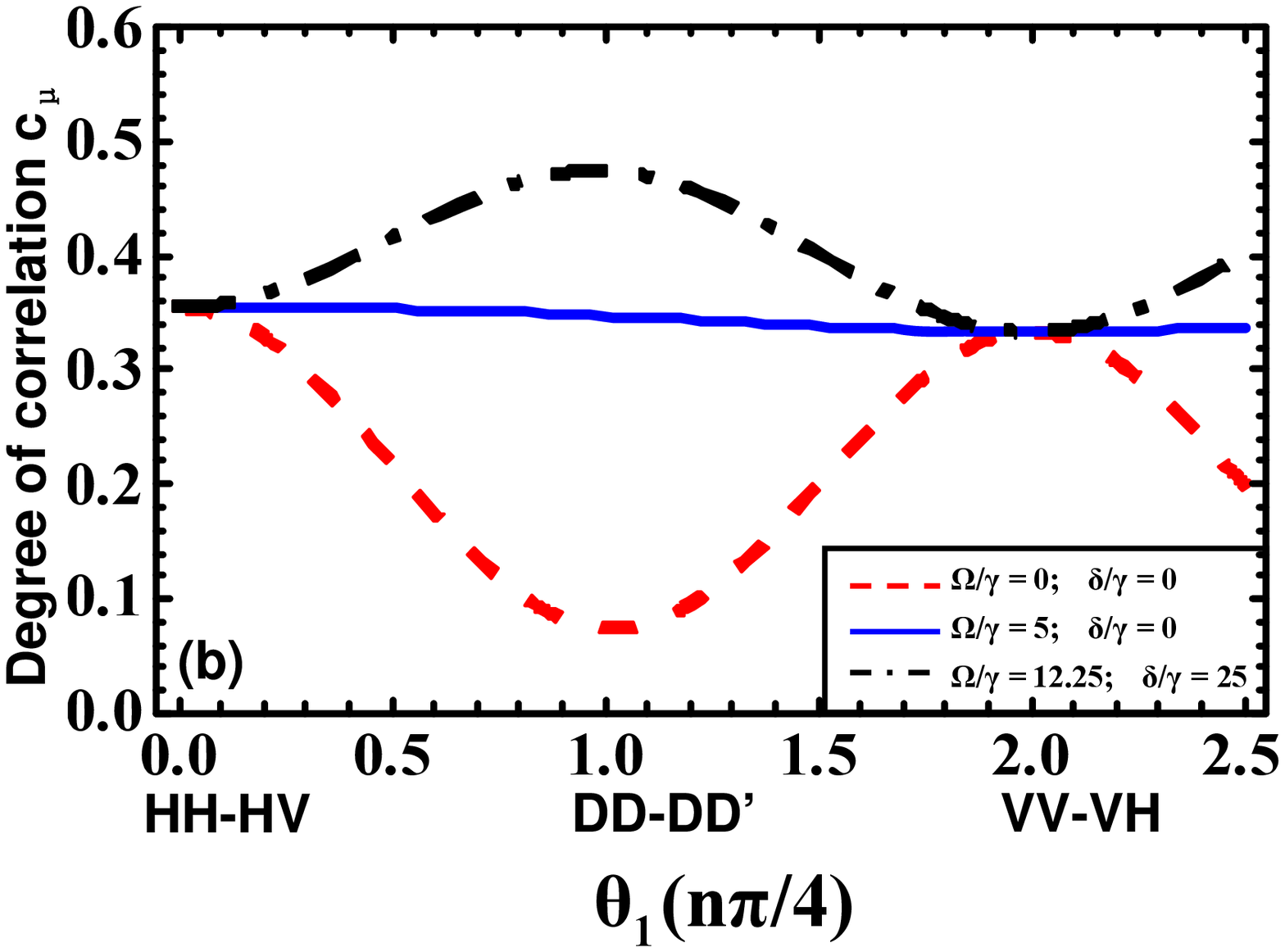}
\caption{(Color online) Degree of correlation averaged over time
as a function of basis angle for different values of the external
field. Here figure (a) correspond to $\Delta_{FS} / \gamma = 0$
and figure (b) is for $\Delta_{FS} / \gamma = 5$. The incoherent
dephasing rate $\gamma_{d}$ of the intermediate level is taken to
be $\gamma_{d} / \gamma = 1$.} \label{AvCorrDephase}
\end{figure}

As discussed earlier, now we numerically investigate the
competition between the coherent (external driving field) and
incoherent process (dephasing of the intermediate levels
$\gamma_{d} \neq 0$) and their effect on the degree of
polarization correlation. Fig. \ref{AvCorrDephase}a \& b shows the
time averaged degree of correlation $C_{\mu}$ for $\Delta_{FS} =
0$ and $\Delta_{FS} = 5\gamma$ respective, in presence of
incoherent dephasing $\gamma_{d} = \gamma$ and different external
fields $\Omega$.  Note that the adverse effect of such incoherent
dephasing on the quantum correlation is clearly visible in both
Figs.(\ref{AvCorrDephase}a \& b). In absence of the external
field, in addition to the fact that $C_{\mu}$ oscillates as the
polarization basis is changed (suggesting classical correlation)
the maximum value of it is reduced ($< 1$). This implies that in
presence of the incoherent process perfect correlation among the
photons is not possible for example in the rectilinear basis also
(see Fig.\ref{AvCorrDephase} a) even when $\Delta_{FS} = 0$. This
is in striking contrast to the case discussed earlier (for
$\Delta_{FS} = 0$) when we have not considered the incoherent
dephasing.

Moreover we see in Fig.(\ref{AvCorrDephase}a) that when
$\gamma_{d} \neq 0, \Delta_{FS} = 0$, in presence of a resonant
external field the behavior of $C_{\mu}$ changes significantly.
The otherwise oscillatory $C_{\mu}$ is now
suppressed and in particular we find that $C_{\mu}$ beomes independent of
polarization basis for $\Omega = \gamma_{d}$. We find that the
difference in the degree of correlation of the photon pairs
between the rectilinear and diagonal basis is reduced by almost
$100\%$. However if we consider fields stronger than the
incoherent dephasing rate the quantum correlation is spoiled again
(as shown by the oscillatory dashed-dot curve in
Fig.\ref{AvCorrDephase}a). Our simulation suggest that for optimal
parameters even in presence of degrading incoherent processes the
coherent field can revive the quantum correlations among the
photons which thereby makes the degree of correlation independent
of the polarization basis. In Fig.(\ref{AvCorrDephase}b) we
consider the competition between the coherent field and incoherent
dephasing process further, but now in presence of a large
excitonic-level splitting. We see similar behavior of the
$C_{\mu}$ as in Fig.(\ref{AvCorrDephase}a) with one main
difference. In this case we see that the external field can revive
the quantum correlation when it is resonant and $\Omega =
\Delta_{FS}$. This is different from the analysis of the earlier
section where $\gamma_{d} =0$, and the optimal condition was found
to be with a non-resonant field. Thus we see from our simulations
that in addition to intermediate level splitting if we have a
incoherent process (dephasing in this case) the quantum
correlation among the photons can be preserved to some extent by
the external field even though the degree of correlation achieved
in this case is not perfect.

\section{Bell's Inequality for Correlated Polarized Photons }
In a classic paper in $1969$ Clauser, Horne, Shimony and Holt \cite{Cla69} re-formulated and generalized Bell's inequality in terms of practically feasible correlation measurements among any two quantum mechanical systems. This later came to be known as \textit{CHSH inequality} and was first measured by Aspect and co-workers in a beautifully designed experiment \cite{Asp82}. In recent times though CHSH inequality has been exploited extensively in studying entanglement among photons \cite{Lar08, You09, Mar04, Moe04} and there application to different quantum information protocols \cite{Eke91}. The standard procedure to verify the CHSH version of Bell�s inequality  \cite{Cla69} for photon polarization states, is as follows. Two independent polarization detectors perform a coincidence measurement on the two photons emitted by the source for four combinations of linear-polarizer angles: Detector $1$ say measures at some angles $\alpha_{1}$ and  $\alpha_{2}$, and detector $2$ at  $\beta_{1}$ and  $\beta_{2}$. The Bell parameter $S$ then calculated in the CHSH form is given by,
\begin{equation}
\label{B1}
S = E( \alpha_{1},  \beta_{1})-E( \alpha_{1},  \beta_{2}) + E( \alpha_{2},  \beta_{1}) + E( \alpha_{2},\beta_{2}) \leq 2,
\end{equation}
where
\begin{equation}
\label{B2}
E(\alpha_{ i},  \beta_{j}) = p_{+}( \alpha_{i},  \beta_{j})-p_{-}(\alpha_{ i}, \beta_{j})
\end{equation}
is the correlation coefficient of the measurement ${\alpha_{i}, \beta_{j}}$.
Here $p_{+}(\alpha_{i}, \beta_{j})$ denotes the fraction of events where the
polarization measurements by detector $1$ at angle $\alpha_{i}$ and by
detector 2 at $\beta_{j}$ are positively correlated (both photons pass
through their respective polarizers, or both are rejected)
and $p_{-}(\alpha_{i},  \beta_{j})$ denotes the fraction of events where the
photons are anticorrelated (one passes the polarizer, and
the other is rejected). If the photons are perfectly correlated,
then $E( \alpha_{i}, \beta_{j}) = +1$; for perfectly anticorrelated
photons, we have $E( \alpha_{i},  \beta_{j}) = -1$. Note that the maximum magnitude of the Bell parameter $S$ that quantum mechanics allows is $|S| = 2\sqrt{2}$ and the states that satisfy this are known as the Bell states (For example Eq. \ref{1}). For the photons generated from the radiative cascade we follow an approach outlined in \cite{Tan08,You09} to define the Bell parameter of (\ref{B1}) in the rectilinear-diagonal polarization basis as,
\begin{equation}
\label{B3}
S = \sqrt{2}\left [C_{H}+C_{D}\right ] \leq 2
\end{equation}
where $C_{H}$ and $C_{D}$ corresponds to the degree of polarization defined by (\ref{14}) in the rectilinear and diagonal basis. Thus if the radiatively emitted photon pairs are correlated following the laws of quantum mechanics we will expect that the above inequality will be violated. However as the correlations are sensitive to the intermediate level splitting and any incoherent mechanisms present in the system we intuitively expect the CHSH inequality to be also sensitive to such system parameters. Thus, it is worth investigating the inequality of Eq. (\ref{B3}) as a function of the intermediate level splitting and the incoherent dephasing rate. Further, we have seen that external field induced A.C. stark shift can diminish the intermediate level splitting in the radiative cascade, thereby reviving the lost quantum correlation among the photons. This hence raises the question as to how does the Bell parameter $S$ behave in presence of a strong external driving field. We next study Eq. (\ref{B3}) in context to these effects and discuss their implications.
\begin{figure}[!h]
\label{bell1}
\includegraphics[width=8.0cm]{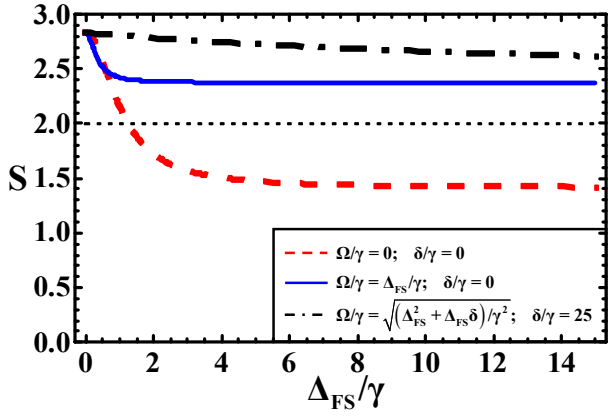}
\caption{(Color online) The Bell parameter as a function of the
intermediate-level splitting $\Delta_{FS}$ in presence and absence
of the external field $\Omega$. No incoherent dephasing has been
considered here. All parameters are normalized with respect to
radiative decay rate $\gamma$.}
\end{figure}

In Fig. 5 we plot the Bell parameter $S$ as a function of the intermediate level splitting $\Delta_{FS}$. We first consider the case with no external field, which is depicted by the broken line in Fig. 5. We find that for the two photon radiative cascade with $\Delta_{FS} = 0$, the violation is maximum. We remind the reader that, this corresponds to the condition when the emitted photons are indistinguishable in  time-frequency and generates an entangled state in the linear polarization basis $\{H, V\}$ (Eq.\ref{1}). However we find that as $\Delta_{FS}$ increases, $S$ decreases and there is no more violations for $\Delta_{FS} > \gamma~(S < 2) $. These behavior of the Bell parameter suggest that for $\Delta_{FS} \neq 0$, but less than the line-width $\gamma$ of the intermediate levels the photons are still quantum mechanically correlated \cite{Mul09}. Further, as the splitting increases beyond $4\gamma$ the Bell parameter becomes a constant at a value of around $S =1.5$. Thus we see that for $\Delta_{FS} > \gamma$ the correlations of photons emitted in the radiative cascade remain no more quantum.

However when the external field is turned on it effects the correlation among the photons dramatically and that is reflected on the Bell parameter. The solid and dashed-dot curve in Fig. 5 shows the Bell parameter as a function of the intermediate level splitting $\Delta_{FS}$ in presence of a resonant and non-resonant field respectively. We see from the behavior of the solid curve that eventhough in presence of a resonant external field $S$ decreases initially with increase in $\Delta_{FS}$ but it never decrease below $2$. Rather it becomes a constant (around $2.5$) as the intermediate level splitting increases and the external field strength is varied to keep it tuned to the level splitting ($\Omega = \Delta_{FS}$). Moreover, for a non-resonant external field (the dashed-dot curve) the effect on $S$ is even more pronounced. This is expected, as from our earlier results, we know that the degree of correlations among the photons are even better when we have detuned external field. In this case we find that with increase in $\Delta_{FS}$ given that $\Omega$ is kept equal to the level splitting, $S$ change minutely from the value at maximum violation.
\begin{figure}[h]
\label{bell2}
\includegraphics[width=8.0cm]{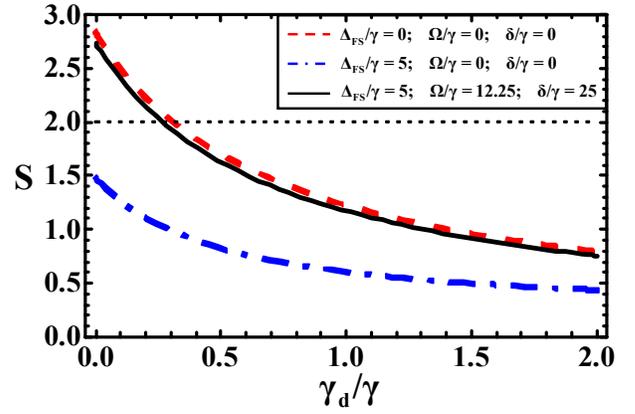}
\caption{(Color online) The Bell parameter as a function of the
incoherent dephasing rate$\gamma_{d}$ for different values of the
external field. Here the intermediate level are considered to be
energy separated by an amount $\Delta_{FS}/\gamma = 5$.}
\end{figure}
In Fig. 6 we plot the Bell parameter $S$ as a function of the incoherent dephasing rate $\gamma_{d}$. We find that when $\Delta_{FS} = 0$ the Bell parameter $S \ge 2$ only for $\gamma_{d}$ much smaller than the linewidth $\gamma$ of the intermediate levels. Further as for non-zero $\Delta_{FS}$ the quantum correlation among the photons are adversely effected we have no violation of Bell's inequality even for $\gamma_{d} = 0$. However by turning on the far detuned external field we see from Fig. 6 that we are successful in regaining the situation for $\Delta_{FS} = 0$ and thus have violation of Bell's inequality for small $\gamma_{d}$. Thus our analysis shows that, even though the external field is very effective in reducing or getting rid of the adverse effect of intermediate level splitting on the quantum correlation of photons, it does not substantially alter or inhibit the effect of decoherence created in the system by the incoherent dephasing.

\section{Conclusion}
In conclusion, we in this paper studied the polarization dependent intensity-intensity correlation of a pair of photons emitted in a four level radiative cascade driven by an external field. We found that, by applying a far detuned external field, the intensity-intensity correlation which is substantially degraded in the presence of the level splitting between the intermediate levels, can be efficiently revived. The mechanism that leads to this revival was found to be an induced stark shift and the formation of dressed states in the system by the non-resonant external field. We further investigated the interplay of  the intermediate level splitting and the external field in presence of a incoherent dephasing of the intermediate levels. The incoherent dephasing create a decoherence in the system and thereby substantially effect the degree of correlation of the photons. In the presence of an external field, however we found that the effect can be partially controlled. Finally, we also investigate the non-locality of the correlations by studying the violation of Bell's inequality in the linear polarization basis for the radiative cascade. For an intermediate level, energy splitting more than the radiative linewidth and in presence of incoherent dephasing rate we found that the photons are classically correlated and there is no violation of Bell's inequality. In presence of an external field and no incoherent processes, effect of the intermediate level splitting can be suppressed, thereby generating nonlocal correlation among the photons. This hence leads to violation of Bell's inequality in the radiative cascade for even arbitrary intermediate level splitting. In presence of incoherent dephasing however, the external field modulation is effective in preserving the non-locality of the correlation only if the incoherent process is not so strong.

\begin{acknowledgments}
L. Y. and S. D. thanks Marlan O. Scully for helpful discussions
and gratefully acknowledge support from the the Office of Naval
Research. S. D also thanks C. H. Keitel, E.Sete and D. Voronine
for ciritical comments on the manuscript. L. Y. is supported by
the Herman F. Heep and Minnie Belle Heep Texas A$\&$M University
Endowed Fund held/administered by the Texas A$\&$M Foundation.
\end{acknowledgments}

\appendix

\section*{Appendix A - Density Matrix Elements}

\renewcommand{\theequation}{A-\arabic{equation}}
\setcounter{equation}{0}

To calculate the two-time intensity-intensity correlation function the
dynamical evolution of the matrix elements $\rho _{X_{1}X_{1}} \left( {t
+ \tau } \right)$, $\rho _{X_{2}X_{2}} \left( {t + \tau }
\right)$, and $\rho _{X_{1}X_{2}} \left( {t + \tau } \right)$ are required. This can be found by solving the equation of motion for the density matrix elements given by the set of equations:
\begin{eqnarray}
\dot \rho _{2X X_{1}}  =  - \frac{1}{2}\left( {\gamma _1  + \gamma
_2 + \gamma _3  + \gamma _{21} } \right)\rho _{2X X_{1}} ;
\end{eqnarray}
\begin{eqnarray}
\begin{array}{l}
 \dot \rho _{2X X_{2}}  =  - \frac{1}{2}\left( {\gamma _1  + \gamma _2  + \gamma _4  + \gamma _u  + \gamma _{12} } \right)\rho _{2X X_{2}}  \\
 \begin{array}{*{20}c}
   {} & {} & {} & {} & {} & { - i\Omega ^* \rho _{2Xu} } ; \\
\end{array} \\
 \end{array}
\end{eqnarray}
\begin{eqnarray}
\dot \rho _{2Xg}  =  - \frac{1}{2}\left( {\gamma _1  + \gamma _2 }
\right)\rho _{2Xg} ;
\end{eqnarray}
\begin{eqnarray}
\dot \rho _{2Xu}  =  - \frac{1}{2}\left( {\gamma _1  + \gamma _2
+i 2 \delta } \right)\rho _{2Xu}  - i\Omega \rho _{2X X_{2}} ;
\end{eqnarray}
\begin{eqnarray}
\dot \rho _{X_{1} g}  =  - \frac{1}{2}\left( {\gamma _3  + \gamma
_{21} } \right)\rho _{X_{1} g} ;
\end{eqnarray}
\begin{eqnarray}
\dot \rho _{X_{1} u}  =  - \frac{1}{2}\left( {\gamma _3  + \gamma
_{21} + i 2 \delta} \right)\rho _{X_{1} u}  - i \Omega \rho
_{X_{1}X_{2}} ;
\end{eqnarray}
\begin{eqnarray}
\begin{array}{l}
 \dot \rho _{X_{1}X_{2}}  =  - \frac{1}{2}\left( {\gamma _3  + \gamma _4  + \gamma _u  + \gamma _{21}  + \gamma _{12} } \right)\rho _{X_{1}X_{2}}  \\
 \begin{array}{*{20}c}
   {} & {} & {} & {} & {} & { - i\Omega ^* \rho _{X_{1} u} } ; \\
\end{array} \\
 \end{array}
\end{eqnarray}
\begin{eqnarray}
\dot \rho _{X_{2} g}  =  - \frac{1}{2}\left( {\gamma _4  + \gamma
_u  + \gamma _{12} } \right)\rho _{X_{2} g}  + i\Omega \rho _{ug}
;
\end{eqnarray}
\begin{eqnarray}
\begin{array}{l}
 \dot \rho _{X_{2} u}  =  - \frac{1}{2}\left( {\gamma _4  + \gamma _u  + \gamma _{12} + i 2 \delta} \right)\rho _{X_{2} u}  \\
 \begin{array}{*{20}c}
   {} & {} & {} & {} & {} & { - i\Omega \left( {\rho _{X_{2}X_{2}}  - \rho _{uu} } \right)} ; \\
\end{array} \\
 \end{array}
\end{eqnarray}
\begin{eqnarray}
\dot \rho _{ug}  = i \delta \rho_{ug}+i\Omega ^* \rho _{X_{2} g}
 ;
\end{eqnarray}
\begin{eqnarray}
\dot \rho _{2X2X}  =  - \left( {\gamma _1  + \gamma _2 }
\right)\rho _{2X2X} ;
\end{eqnarray}
\begin{eqnarray}
\begin{array}{l}
 \dot \rho _{X_{1}X_{1}}  =  - \left( {\gamma _3  + \gamma _{21} }
\right)\rho _{X_{1}X_{1}}  \\
 \begin{array}{*{20}c}
   {} & {} & {} & {} & {} & {  + \gamma _1 \rho _{2X2X}  + \gamma
_{12} \rho _{X_{2}X_{2}}  } ; \\
\end{array} \\
 \end{array}
\end{eqnarray}
\begin{eqnarray}
\begin{array}{l}
 \dot \rho _{X_{2}X_{2}}  =  - \left( {\gamma _4  + \gamma _u  + \gamma _{12} } \right)\rho _{X_{2}X_{2}}  + \gamma _2 \rho _{2X2X}  \\
 \begin{array}{*{20}c}
   {} & {} & {} & {} & {} & { + \gamma _{21} \rho _{X_{1}X_{1}}  + i\Omega \rho _{uX_{2}}  - i\Omega ^* \rho _{X_{2}u} } ; \\
\end{array} \\
 \end{array}
\end{eqnarray}
\begin{eqnarray}
\dot \rho _{uu}  = \gamma _u \rho _{X_{2}X_{2}}  - i\Omega \rho
_{uX_{2}}  + i\Omega ^* \rho _{X_{2}u} ;
\end{eqnarray}

\begin{eqnarray}
\rho_{ji} = \rho^{\ast}_{ij}
\end{eqnarray}

\begin{eqnarray}
\rho _{2X2X}  + \rho _{X_{1}X_{1}}  + \rho _{X_{2}X_{2}} + \rho
_{uu} + \rho _{gg}  = 1 .
\end{eqnarray}

\section*{Appendix B - Correlation function}

\renewcommand{\theequation}{B-\arabic{equation}}
\setcounter{equation}{0}

The resulting two-time correlation function in Eq.
(\ref{6}) can written as
\begin{eqnarray}
\begin{array}{l}
 \left\langle {II} \right\rangle  = \left( {\frac{{\omega _0 }}{c}} \right)^8 \frac{1}{{4r^4 }}D_1^2 D_2^2 \left\langle {\left| 2X \right\rangle \left\langle 2X \right|_t } \right\rangle \left\{ {f_1 \left( \tau  \right) + f_2 \left( \tau  \right)} \right. \\
  + g_1 \left( \tau  \right) + g_2 \left( \tau  \right) + \left( {\cos 2\theta _1  + \cos 2\theta _2 } \right)\left( {f_1 \left( \tau  \right) - g_1 \left( \tau  \right)} \right) \\
  + \left( {\cos 2\theta _1  - \cos 2\theta _2 } \right)\left( {g_2 \left( \tau  \right) - f_2 \left( \tau  \right)} \right) \\
  + \cos 2\theta _1 \cos 2\theta _2 \left( {f_1 \left( \tau  \right) + g_1 \left( \tau  \right) - f_2 \left( \tau  \right) - g_2 \left( \tau  \right)} \right) \\
 \left. { + \sin 2\theta _1 \sin 2\theta _2 \left[ {e^{ - i\left( {\phi _1  + \phi _2 } \right)} w^ *  \left( \tau  \right) + e^{i\left( {\phi _1  + \phi _2 } \right)} w\left( \tau  \right)} \right]} \right\}. \\
 \end{array}\label{correlation}
\end{eqnarray}
Here
\begin{eqnarray}
\begin{array}{l}
 f_1 \left( \tau  \right) = e^{b_0 \tau } \left[ {\cosh \left( {\frac{{\eta \tau }}{2}} \right)} \right. \\
 \begin{array}{*{20}c}
   {} & {} & {} & {} & {} & {\left. { - \frac{{\gamma _3  - \gamma _4  + \gamma _{21}  - \gamma _{12}  - \gamma _u }}{\eta }\sinh \left( {\frac{{\eta \tau }}{2}} \right)} \right]} ; \\
\end{array} \\
 \end{array}
\end{eqnarray}
\begin{eqnarray}
f_2 \left( \tau  \right) = \frac{{2\gamma _{12} e^{b_0 \tau }
}}{\eta }\sinh \left( {\frac{{\eta \tau }}{2}} \right) ;
\end{eqnarray}
\begin{eqnarray}
\begin{array}{l}
 g_1 \left( \tau  \right) = e^{b_0 \tau } \left[ {\cosh \left( {\frac{{\eta \tau }}{2}} \right)} \right. \\
 \begin{array}{*{20}c}
   {} & {} & {} & {} & {} & {\left. { + \frac{{\gamma _3  - \gamma _4  + \gamma _{21}  - \gamma _{12}  - \gamma _u }}{\eta }\sinh \left( {\frac{{\eta \tau }}{2}} \right)} \right]} ; \\
\end{array} \\
 \end{array}
\end{eqnarray}
\begin{eqnarray}
g_2 \left( \tau  \right) = \frac{{2\gamma _{21} e^{b_0 \tau }
}}{\eta }\sinh \left( {\frac{{\eta \tau }}{2}} \right) ;
\end{eqnarray}
\begin{eqnarray}
\begin{array}{l}
 w\left( \tau  \right) = e^{a_0 \tau  - i\Delta_{FS} \tau } \left[ {\cos \left( {\frac{{\mu \tau }}{4}} \right)} \right. \\
 \begin{array}{*{20}c}
   {} & {} & {} & {} & {} & {\left. { + \frac{{\gamma _4  + \gamma _{12}  + \gamma _u  - 2i\delta }}{\mu }\sin \left( {\frac{{\mu \tau }}{4}} \right)} \right]} , \\
\end{array} \\
 \end{array}
\end{eqnarray}
where,~ $a_0  =  - \frac{1}{4}\left( {2\gamma _3  + 2\gamma _{21} +
\gamma _4  + \gamma _{12}  + \gamma _u  + 2i\delta } \right)$,\\
\\
$b_0  =  - \frac{1}{2}\left( {\gamma _3  + \gamma _4  + \gamma
_{21}  + \gamma _{12}  + \gamma _u } \right)$, \\
\\
$\mu  = \sqrt{16\Omega ^2  - \left( {\gamma _4  + \gamma _{12}  + \gamma _u  -
2i\delta } \right)^2 }$, and \\
\\
$\eta  = \sqrt {4\gamma _{21} \gamma
_{12}  - \left( {\gamma _3  - \gamma _4  + \gamma _{21}  - \gamma
_{12}  - \gamma _u } \right)^2 }$. \\
\\
Also, $\mathcal{D}_1  = \left|
{\vec d_{X_{1}2X} } \right| = \left| {\vec d_{X_{2}2X} } \right|$
and $\mathcal{D}_2 = \left| {\vec d_{gX_{1}} } \right|  = \left|
{\vec d_{gX_{2}} } \right|$.

We can calculate the average correlation function as $\left\langle
{II} \right\rangle _{av}  = \int_0^\infty {\left\langle {II}
\right\rangle d\tau }$ and the result is
\begin{eqnarray}
\begin{array}{l}
 \left\langle {II} \right\rangle _{av}  = \left( {\frac{{\omega _0 }}{c}} \right)^8 \frac{1}{{4r^4 }}D_1^2 D_2^2 \left\langle {\left| e \right\rangle \left\langle e \right|_t } \right\rangle \left\{ {F_1  + F_2  + G_1  + G_2 } \right. \\
  + \left( {\cos 2\theta _1  + \cos 2\theta _2 } \right)\left( {F_1  - G_1 } \right) \\
  + \left( {\cos 2\theta _1  - \cos 2\theta _2 } \right)\left( {G_2  - F_2 } \right) \\
  + \cos 2\theta _1 \cos 2\theta _2 \left( {F_1  + G_1  - F_2  - G_2 } \right) \\
 \left. { + \sin 2\theta _1 \sin 2\theta _2 \left[ {e^{ - i\left( {\phi _1  + \phi _2 } \right)} W^ *   + e^{i\left( {\phi _1  + \phi _2 } \right)} W} \right]} \right\} , \\
 \end{array}
\end{eqnarray}
where
\begin{eqnarray}
F_1  = \frac{{\gamma _4  + \gamma _{12}  + \gamma _u }}{{b_0^2  -
\eta ^2 /4}} ;
\end{eqnarray}
\begin{eqnarray}
F_2  = \frac{{\gamma _{12} }}{{b_0^2  - \eta ^2 /4}} ;
\end{eqnarray}
\begin{eqnarray}
G_1  = \frac{{\gamma _3  + \gamma _{21} }}{{b_0^2  - \eta ^2 /4}}
;
\end{eqnarray}
\begin{eqnarray}
G_2  = \frac{{\gamma _{21} }}{{b_0^2  - \eta ^2 /4}} ;
\end{eqnarray}
\begin{eqnarray}
W = \frac{{i\left( {\Delta_{FS}  + \delta } \right) + \left( {\gamma _3
+ \gamma _{21} } \right)/2}}{{\left( {a_0  - i\Delta_{FS} } \right)^2 +
\mu ^2 /16}} .
\end{eqnarray}
For the case $\gamma _3  = \gamma _4  \equiv \gamma$, $\gamma
_{21}  = \gamma _{12}$, and $\gamma ,\gamma _{12}  \gg \gamma _u$,
the two-time polarization-angle-dependent intensity-intensity
correlation function is found to be
\begin{eqnarray}
\begin{array}{l}
 \left\langle {II} \right\rangle  = \left( {\frac{{\omega _0 }}{c}} \right)^8 \frac{1}{{2r^4 }}D_1^2 D_2^2 \left\langle {\left| 2X \right\rangle \left\langle 2X \right|_t } \right\rangle \left\{ {e^{ - \gamma \tau } } \right. \\
  + \cos 2\theta _1 \cos 2\theta _2 e^{ - \left( {\gamma  + 2\gamma _{12} } \right)\tau }  \\
 \left. { + \sin 2\theta _1 \sin 2\theta _2 {\mathop{\rm Re}\nolimits} \left[ {e^{i\left( {\phi _1  + \phi _2 } \right)} w\left( \tau  \right)} \right]} \right\} .\\
 \end{array}
\end{eqnarray}
\\


\begin{thebibliography}{99}
\bibitem{Koc67}
C. A. Kocher and E. D. Commins, Phys. Rev. Lett. {\bf 18}, 575 (1967).

\bibitem{Cla69}
J. F. Clauser, M. A. Horne, A. Shimony, and R. A. Holt, Phys. Rev. Lett {\bf 23}, 880 (1969).

\bibitem{Fre72}
S. J. Freedman and J. F. Clauser,  Phys. Rev. Lett.  {\bf 28}, 938 (1972).

\bibitem{Fry76}
E. S. Fry and R. C. Thompson, Phys. Rev. Lett. {\bf 37}, 465 (1976).

\bibitem{Asp82}
A. Aspect, P. Grangier, and G. Roger, Phys. Rev. Lett. {\bf 49}, 91 (1982).

\bibitem{Bel65}
J. S. Bell, Physics (Long Island City, N. Y.) {\bf 1}, 195 (1965).

\bibitem{Eke91}
A. K. Ekert, Phys. Rev. Lett. {\bf 67}, 661 (1991).

\bibitem{Urs04}
R. Ursin, T. Jennewein, M. Aspelmeyer, R. Kaltenbaek, M.
Lindenthal, and A. Zeilinger, Nature  London  {\bf 430}, 849
(2004) .

\bibitem{Lan07}
O. Landry, J. A. W. van Houwelingen, A. Beveratos, H. Zbinden,
and N. Gisin, J. Opt. Soc. Am. B {\bf 24},  398 (2007) .

\bibitem{Kni01}
E. Knill, R. Laflamme and G. J. Milburn,  Nature {\bf 409}, 46 (2001).

\bibitem{Bre98}
H-J. Briegel, W. Dur, J. I. Cirac and P. Zoller, Phys. Rev. Lett. {\bf 81}, 5932 (1998).

\bibitem{Kie93}
T. E. Kiess, Y. H. Shih, A.V. Sergienko, and C. O. Alley,
Phys. Rev. Lett.{\bf 71}, 3893 (1993).

\bibitem{Kwi95}
P. G. Kwiat \textit{et al}., Phys. Rev. Lett. {\bf 75}, 4337 (1995).

\bibitem{Sca05}
V. Scarani, H. de Riedmatten, I. Marcikic, H. Zbinden, and N.
Gisin, Eur. Phys. J. D {\bf 32}, 129 (2005).

\bibitem{Lar08}
M. Larqu\'{e}, I. Robert-Philip, and A. Beveratos,
\emph{Phys. Rev. A.} {\bf 77}, 042118 (2008).

\bibitem{Das08}
Sumanta Das and G.S. Agarwal, \emph{J. Phys. B:
At. Mol. Opt. Phys.} {\bf 41}, 225502 (2008).

\bibitem{Ben00}
O. Benson, C. Santori, M. Pelton, and Y. Yamamoto, Phys.
Rev. Lett. {\bf 84}, 2513 (2000).

\bibitem{Sim05}
C. Simon and J. P. Poizat, Phys. Rev. Lett. {\bf 94}, 030502 (2005).

\bibitem{Ste_n06}
R. M. Stevenson, R. J. Young, P. Atkinson, K. Cooper,
D. A. Ritchie, and A. J. Shields, Nature (London) {\bf 439},
179 (2006).

\bibitem{Ako06}
N. Akopian, N. H. Lindner, E. Poem, Y. Berlatzky, J.
Avron, and D. Gershoni, Phys. Rev. Lett. {\bf 96}, 130501
(2006).

\bibitem{Haf07}
R. Hafenbrak, S. M. Ulrich, P. Michler, L. Wang, A.
Rastelli, and O. G. Schmidt, New J. Phys. {\bf 9}, 315 (2007).

\bibitem{Gam96}
D. Gammon, E. S. Snow, B. V. Shanabrook, D. S. Katzer, and
D. Park, Phys. Rev. Lett. {\bf 76}, 3005 (1996) .

\bibitem{Bay02}
M. Bayer, \textit{et al}. Phys. Rev. B {\bf 65}, 195315 (2002) .

\bibitem{San02}
C. Santori,\textit{et. al}. Phys. Rev. B {\bf 66}, 045308 (2002).

\bibitem{Seg05}
R. Seguin, S. Schliwa, S. Rodt, K. Po�tschke, U.W. Pohl,
and D. Bimberg, Phys. Rev. Lett. {\bf 95}, 257402 (2005).

\bibitem{Ste06}
R. M. Stevenson, R. J. Young, P. See, D. G. Gevaux, K.
Cooper, P. Atkinson, I. Farrer, D.A. Ritchie, and A. J.
Shields, Phys. Rev. B {\bf 73}, 033306 (2006).

\bibitem{Ger07}
B. D. Gerardot, S. Seidl, P. A. Daigarno, R. J. Warburton,
D. Granados, J. M. Garcia, K. Kowalik, and O. Krebs,
Appl. Phys. Lett. {\bf 90}, 041101 (2007).

\bibitem{Jun08}
Gregor Jundt \textit{et. al}. Phys. Rev. Lett. {\bf 100}, 177401 (2008).

\bibitem{Mul09}
A. Muller, W. Fang, J. Lawall, and G. S. Solomon, Phys. Rev. Lett. {\bf 103}, 217402 (2009).

\bibitem{Ber06}
A. Berthelot, I. Favero, G. Cassabois, C. Voisin, C. Delalande,
Ph. Roussignol, R. Ferreira, and J. M. G�rard, Nat. Phys. {\bf 2},
759 (2006) .

\bibitem{You09} R.J. Young, R.M. Stevenson, A.J. Hudson,
C.A. Nicoll, D.A. Ritchie, and A.J. Shields, \emph{Phys. Rev.
Lett.} {\bf 102}, 030406 (2009).

\bibitem{Mar04}
I. Marcikic, H. de Riedmatten, W. Tittel, H. Zbinden,
M. Legre, and N. Gisin, Phys. Rev. Lett. {\bf 93}, 180502
(2004).

\bibitem{Moe04}
D. L. Moehring, M. J. Madsen, B. B. Blinov, and C. Monroe,
Phys. Rev. Lett. {\bf 93}, 090410 (2004).

\bibitem{Tan08}
H. Tanji, \textit{et. al.}, arxiv, (2008).

\end{thebibliography}
\end{document}